\documentclass{llncs}

\usepackage{graphicx}
\usepackage{xcolor}
\usepackage[normalem]{ulem}

\usepackage{lscape}

\usepackage[export]{adjustbox}

\usepackage{float}

\begin{document}

\title{Quality Attributes in Practice:\\ Contemporary Data} 
%Understanding the causal relationship between software process and quality.
\author{Rasul Tumyrkin\inst{1} \and Manuel Mazzara\inst{1} \and Mohammad Kassab\inst{1,2} \\ \and Giancarlo Succi\inst{1} \and JooYoung Lee\inst{1}} 

\institute{Innopolis university, Innopolis, Russia,\\
\email{\{r.tumyrkin,m.mazzara, m.kassab,g.succi,j.lee\}@innopolis.ru}
\and
Pennsylvania State University, Malvern, USA
}

%\date{}

\maketitle

\begin{abstract}
%\sout{Software development process is indivisible from software architecture nowadays. The latter is the foundation for the consistent implementation of the software design to achieve the end purpose of the development.}% * <giancarlo.succi@gmail.com> 2015-12-05T16:10:49.067Z:
%
% > Software development process is indivisible from software architecture nowadays. The latter is the foundation for the consistent implementation of the software design to achieve the end purpose of the development. 
%
% I would eliminate this part, if you do not mind...
%
% ^ <giancarlo.succi@gmail.com> 2015-12-05T17:47:31.103Z.
It is well known that the software process in place impacts the quality of the resulting product. However, the specific way in which this effect occurs is still mostly unknown and reported through anecdotes.  To gather a better understanding of such relationship, a very large survey has been conducted during the last year and has been completed by more than 100 software developers and engineers from 21 countries. We have used the percentage of satisfied customers estimated by the software developers and engineers as the main dependent variable. The results evidence some interesting patterns, like that quality attribute of which customers are more satisfied appears functionality, architectural styles may not have a significant influence on quality, agile methodologies might result in happier customers, larger companies and shorter projects seems to produce better products.

\end{abstract}

\section{Introduction} 
\label{sec:1}
% * <giancarlo.succi@gmail.com> 2015-12-05T17:46:37.546Z:
%
% > Introduction
%
% I think that here we need to define exactly the purpose of the paper, which I do not fully understand yet. What part of the questionnaire are we analysing?
%
% I think that here we need to define exactly the purpose of the paper, which I do not fully understand yet. What part of the questionnaire are we analysing?
%
% ^.
Quality is the set of characteristics of an entity that describe  its ability to satisfy stated and implied needs of the customer and/or of the end user; this notion has been formalized in numerous standards like the ISO 9000. For systems including software, the notion of quality has been instantiated in several standards, including the ISO 9216 \cite{ISOIEC9126} and IEEE 730 \cite{6835311}. Software Quality is an essential and distinguishing attribute of the final product. Nevertheless, functionality usually takes the front seat during software development. This is mainly because of the nature of these quality requirements which poses a challenge when taking the choice of treating them earlier in the software development. Quality requirements are subjective, relative and they become scattered among multiple modules when they are mapped from the requirements domain to the solution space.  Furthermore, Quality requirements can often interact, in the sense that attempts to achieve one can help or hinder the achievement of the other at particular software functionality. Such an interaction creates an extensive network of interdependencies and tradeoffs among quality requirements which is not easy to trace or estimate \cite{Chung}.

This preference for functionality over the qualities is shortsighted though. Software systems are often redesigned not because they are functionally deficient but because they are too slow, not user friendly, hard to scale or hard to maintain. In addition, quality requirements drive architectural structure more than functionality. In fact, if functionality were the only thing that mattered, there wouldn’t be a need to divide the system into architectural components at all; a single monolithic blob with no internal structure would satisfy the need \cite{Garlan}.

The key issue in implementing an improvement in industrial practices is to first identify the areas that need the most improvement. But little contemporary data exists to report on how the quality requirements and its tight coupling to architecture are perceived in industry. To remedy this deficiency and provide useful data to other researchers we conducted an exploratory survey study on quality requirements and software architecture in practice. In this article, we report on our findings from this survey. Reported data includes characteristics of projects, practices, organizations, and practitioners linked to projects’ qualities and their architectural structures.

While there is an endless list of qualities a software system may have to exhibit, the focus of this survey is on the pre-dominant ones as described in \cite{SA}: Availability, Interoperability, Modifiability, Performance, Security, Testability and Usability.

The rest of this paper is organized as follows: section \ref{sec:2} describes tools and techniques to build questions and collect answers; section \ref{sec:3} reports on the nature of respondents profiles and the businesses in which they are employed; section \ref{sec:4} shows the actual empyrical data and section \ref{sec:5} draws some preliminary conclusions that have to be validate in future with the collection of further data and expansion of dataset.

%\section{Related works [MM+MK]}

\section{Experimental Design}
\label{sec:2}
A web-based survey instrument was created using the web-based QuestionPro survey tool (www.QuestionPro.com). The survey consisted of 19 questions. The survey questions were designed after a careful review to specialized literature on conducting survey studies (e.g. \cite{Hoinville:1982}, \cite{Silverman:2000}, \cite{Marshall:1989}, \cite{methodology1}, \cite{methodology2}). A summary of our survey questions is available via the link\footnote[1]{http://www.questionpro.com/a/summaryReport.do?surveyID=4182537}.
 While the respondents reported a wide range of experiences; they were asked to base their responses on only one software project that they were either currently involved with or had taken part in during the past five years. 
 
Using the conjectures in our hypotheses as means of constructing specific questions, the survey was arranged into five sections: First section aiming at capturing general project characteristics first. Then, a series of questions were asked in the second section to determine the participants’ knowledge of architectural styles and whether if any were applied into the surveyed projects. In case of incorporating architectural styles into the projects; the respondents were then asked  to report on the criteria they used to select these styles in the third section and the challenges they faced while incorporating them in the forth section.
Since quality requirements are the major that shapes the software architecture \cite{SA}; a series of questions were then asked in the fifth section to report on the level of customer’s satisfaction with these qualities while the final product is in use. 

We drew our survey participants from multiple sources but primarily from  members of the following Linked-In professional groups, to which one or more of the authors belonged: ``Software Engineering Productivity: Software Architecture", ``Techpost Media",``ISMG: Software Architecture" and ``ISMG Architecture World". A invitation on these groups was posted under the subject ``Software architecture in practice". The participation to this survey was entirely anonymous and voluntarily. Survey data was collected from May 2015 through September 2015. The survey drew 687 participants from 37 countries. Of these survey takers; 103 completed the survey to the end. The completion rate was 15\%\ and the average time taken to complete the survey was 10 minutes. We also included the results of the partially completed responses. When respondents aborted the survey, they tended to do so on or near question 15, we speculate from “survey fatigue.” 

\section{Profiles of the Respondents}
\label{sec:3}
 In this analysis we take into consideration also the answers given by people who partially completed the questionnaire for statistical analysis not requiring pairing or correlating information, as suggested in \cite{Kassab}. Given this population, responses to the survey are more likely to reflect the opinions and biases of any given project’s development team rather than those of other groups represented in a software development effort. In this section we will consider two aspects of respondents' profiles: \emph{distribution of business} and \emph{type of respondents}.
\subsubsection*{Distribution of business}The distribution of businesses that survey respondents have associated themselves with entails a lot of different fields. The data indicate that respondents are well distributed across a wide range of business domains. All fourteen of the provided domains have been selected at least by few participants. Furthermore, the \textit{``other"} category included responses such as social media, transportation, automotive, virtualisation, meteo and etc.
%(see see Figure ~\ref{fig:domains})

%\begin{figure}[H]
%\center{\includegraphics[scale=0.53, frame]{fields}}
%\caption{Which of the following application domains does/did this project apply to? }
%\label{fig:domains}
%\end{figure}

\subsubsection*{Type of respondents} In order to understand the types of respondents, a number of questions regarding \textit{``Organizational Characteristics"} have been asked.
 %As the data in Figure ~\ref{fig:position} illustrates, 
 Respondents to this survey characterize themselves as programmers and developers 41\% of the time, and software engineers, 17\% percent of the time. One third of the respondents characterize themselves as architects and 9\% percent as managers (project managers, scrum managers and product owners).Other respondents include system engineers, testers, consultants - reaching a total of around 3\%.The majority (more than 36\%) of respondents represent small companies, with an annual budget of less than 5 million US dollars, within the listed business domains (all company sizes are measured both, in terms of annual budget and number of employees). It is noticeable that about one third of respondents ignore  the budget of their companies (this is consistent with the fact that developers often are not exposed to financial information).

%\begin{figure}[H]
%\center{\includegraphics[scale=0.4,frame]{position}}
%\caption{Which of the following describes your major position while engaged in this project?}
%\label{fig:position}
%\end{figure}

%Figure ~\ref{fig:budget} illustrates how 

%\begin{figure}[H]
%\center{\includegraphics[scale=0.5, frame]{budget}}
%\caption{What is/was the approximate size of your organization's annual budget?}
%\label{fig:budget}
%\end{figure}

\section{Analysis of the Results}
\label{sec:4}
%\section{Quality Attributes: State of Practice [MM]}
In this section we will present a portion of the collected data with some projections in order to identify and highlight some aspects of Quality Attributes and satisfaction.
The collected data allow several projections and analysis that cannot be reported in full in this paper. due to space constraints. In Section \ref{sec:5} we will discuss these results in more detail and we will try to connect the dots. We will also anticipate how the work can be continued. In this section in particuar, and mor ein general in this paper, we focus on Quality Attributes and satisfaction.  All data are  based on the responses of IT-specialists and we use the ``satisfaction rate'' as the main measure in evaluating quality of final product. The ``satisfaction rate'' is defined as the estimation made by IT-specialists of the percentage of customers satisfied with quality attributes and overall quality of the final product.

%HERE WE NEED TO PUT SOME COMMENTS ON SUCH MEASURE AND ITS REPRESENTATIVENESS

\begin{figure}[h]
\center{\includegraphics[scale=0.58, frame]{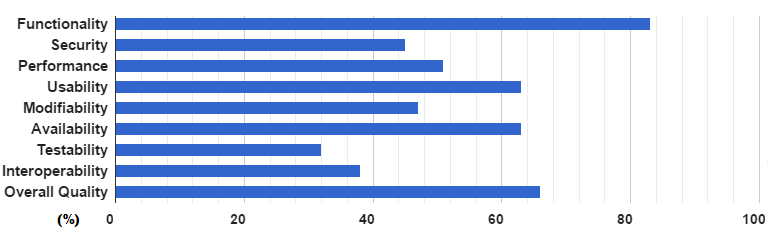}}
\caption{The satisfaction level of the customer with this project in terms of QA}
\label{fig:general_satisfaction}
\end{figure}

Figure ~\ref{fig:general_satisfaction} shows the overall satisfaction rate for each specific Quality Attribute considered in the survey. In this figure over hundreds responses of the survey have been computed. Functionality appears as the QA for which customers have an overall higher satisfaction level with over 80\% of customers satisfied with functionality of the final product. Availability, usability and overall quality also shows high level of satisfaction while security, performance and modifiability suited only about half of customers. As for testability and interoperability, just one out of three respondents reported customers' satisfaction with respect to these attributes. These results can be interpreted according to a specific attention of software engineers on on functionality, usability and availability resulting in higher overall quality of the final software artifacts.    

\begin{figure}[H]
\center{\includegraphics[scale=0.55, frame]{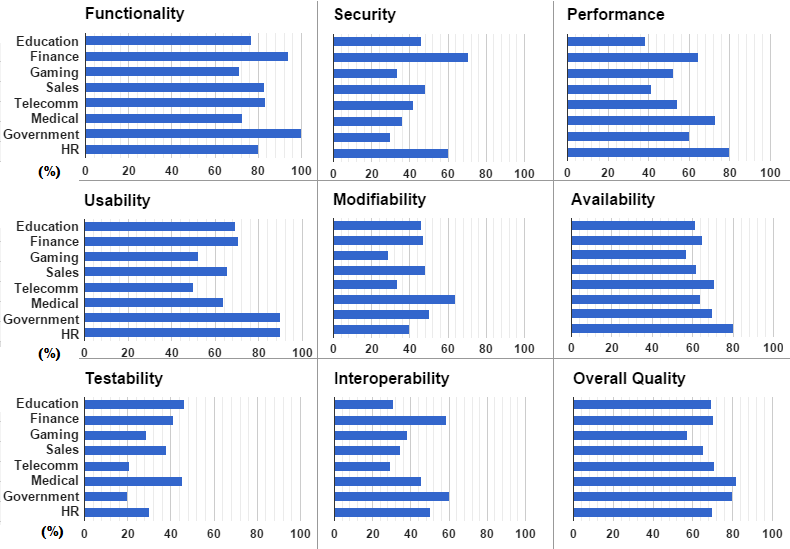}}
\caption{Satisfaction level of the customer with QA for different industry domains}
\label{fig:domain_satisfaction}
\end{figure}

Figure ~\ref{fig:domain_satisfaction} considers Quality Attributes satisfaction rates for specific industry domains. Here only fields which had ten or more responses have been computed, therefore eight industry domains are shown: Education, Finance \& Banking, Gaming \& Utilities, Sales \& Business Development, Telecommunications, Medical Systems \& Pharmaceuticals, Government, Human Resources \& Payroll. The level of customer satisfaction has been here computed individually for each domain.

Regarding \textit{Functionality} each industry domain reports level of satisfactions of 75\% or higher with Finance \& Banking and Government projects almost reaching 100\%. As for \textit{Security}, Government projects shows an unexpectedly low 30\% not very far from games and utilities. Here HR and Finance \& Banking fields reports the highest level. The highest satisfaction levels with \textit{Performance} are shown in HR \& Payroll systems and Medical systems, while Education and Sales scores the least.

\textit{Usability} satisfaction rates have Government and HR projects around 90\% and Gaming and Telecommunication with 50\%. When it comes to \textit{Modifiability}  for Gaming \& Utilities also have very low 30\% levels and Medical Systems score the highest (more than 60\%). More than half customers met the expectations with respect to \textit{Availability} in all business domain. In general, customers have not been much satisfied by \textit{Testability}, where percentages stay between 45\% and 20\%. \textit{Interoperability} also shows low average satisfction levels. Satisfaction concerning \textit{Overall Quality} of the final product is generally high, in particular for Medical Systems and Government projects. 

Summing up data from Figure~\ref{fig:domain_satisfaction} we can conclude that satisfaction rates appear coherent with expectations of a business domain for a specific quality attribute, for example Security is a priority for Finance \& Banking sector. General low rates appearing for Gaming \& Utilities can be explained by the peculiar nature and needs of this domain.

\begin{figure}[h]
\center{\includegraphics[scale=0.55, frame]{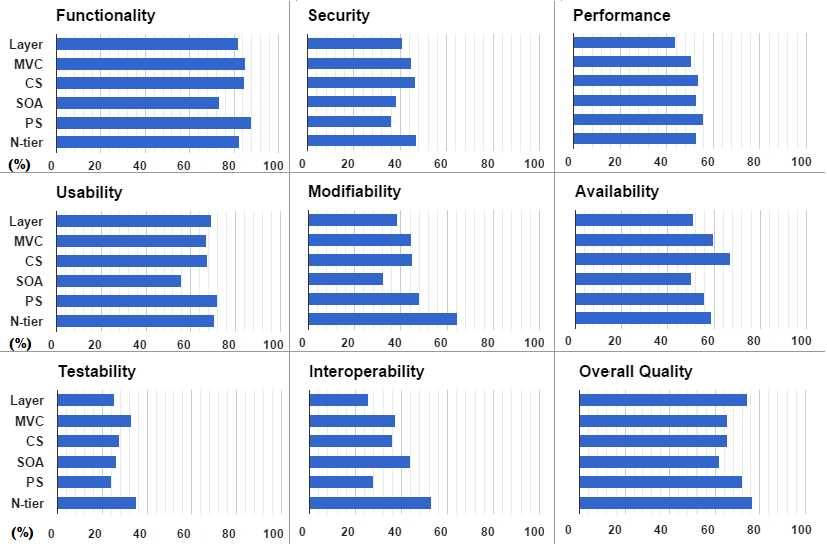}}
\caption{Satisfaction level of the customer with QA for different architectural styles}
\label{fig:arch_satisfaction}
\end{figure}

Figure ~\ref{fig:arch_satisfaction} shows the impact of architectural styles on the final product. Levels of satisfaction related to six among the most popular architectural styles have been projected. Layer, Publisher-Subscriber and Multi-tier are generally higher than MVC, SOA and Client server. However, no significant difference can be appreciated among different architectures used in projects and the fact that architectural style do not affect quality of final product can be concluded.

\begin{figure}[h]
\center{\includegraphics[scale=0.63, frame]{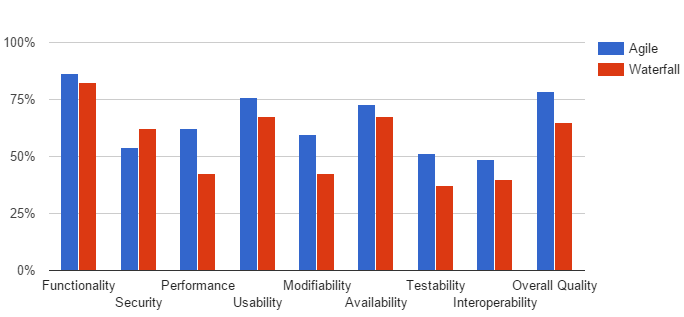}}
\caption{Satisfaction level of the customer with QA for Agile vs. Waterfall projects}
\label{fig:methods_satisfaction}
\end{figure}

We have also analyzed quality attributes from the point of view of software development methods, and observed how projects performed following Agile methodologies shows higher rates of satisfaction than those following Waterfall  (see Figure ~\ref{fig:methods_satisfaction}). %This data contradicts the general belief that quality is not affected by development methods, in particular .... \textbf{TODO: REPORT LITERATURE TO THIS REGARD: MK}

\begin{figure}[h]
\center{\includegraphics[scale=0.5, frame]{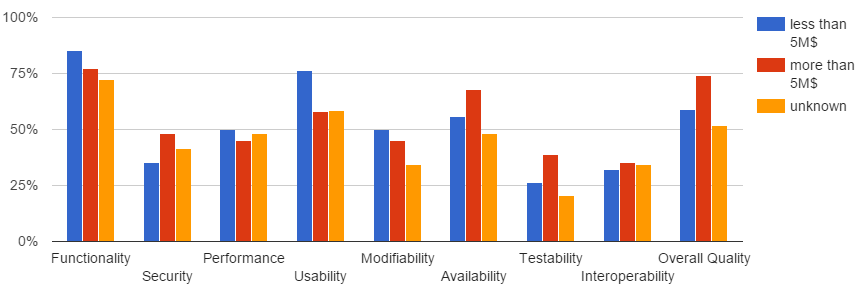}}
\caption{Satisfaction level of the customer with QA for different annual budget}
\label{fig:budget_satisfaction}
\end{figure}

Figure ~\ref{fig:budget_satisfaction} we observe how the overall quality of final products developed by companies with an annual budget higher than 5M of US dollars is higher. This could be considered as the result of the fact that large companies might have more slack time for developers \cite{DeMarco:2002}, who can then devote more time to refactoring or to improve their own development skills.

\begin{figure}[h]
\center{\includegraphics[scale=0.53, frame]{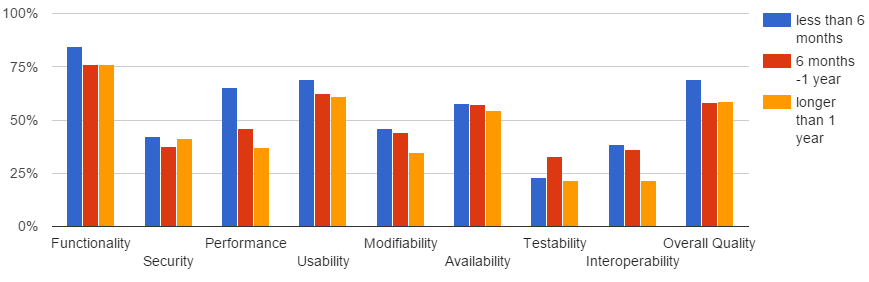}}
\caption{Quality Attributes and project duration}
\label{fig:duration_satisfaction}
\end{figure}

Figure~\ref{fig:duration_satisfaction} shows that projects with a duration inferior to six months have satisfaction rates higher than projects with a longer duration. This can be interpreted along the line that longer development activities result in more complex projects for which may be harder to maintain high quality and more generally keep customer's satisfaction high for all the process.

%\textbf{TODO: ADD A SUMMARY TABLE IN THE LONGER VERSION}

\section{Conclusions}
\label{sec:5}
In the current literature there is no wealth of empirical data on Quality Attributes related to industrial projects. In this survey, data from more than 100 software developers and engineers have been collected and the results on Quality Attributes and satisfaction levels have been reported. In this section we report a synthesis of the findings.

\subsubsection*{Overall satisfaction} The quality attribute of which customers are \textbf{more satisfied} is \emph{functionality}. General positives scores also appears for \emph{availability}, \emph{usability} and \emph{overall quality}. On the opposit side \emph{security}, \emph{performance}, and \emph{modifiability} generate concerns.

\subsubsection*{Specificity of business domains}
The satisfaction rates in a specific business domain of the different quality attributes appear coherent with nature of such domain, for example usability score is very high for Government and HR while functionality scores high for telecommunication. On the other side testability is low for all this business domains. This confirms expectations in a sperimental manner.

\subsubsection*{Architectural Styles} While the certain architectural styles (Layer, Publisher-Subscriber, and Multi-tier) result in higher satisfaction rates than other (MVC, SOA, and Client server), no statistically significant difference can be observed, so that there is no evident relationship between architectural style and quality of final product.

\subsubsection*{Development Methods} Projects implemented using \textbf{Agile methodologies} shows higher rates of satisfaction than those following Waterfall.

\subsubsection*{Budget} The overall quality of final products developed by companies with an annual budget higher than 5M of US dollars is higher than those with a lower budget. 

\subsubsection*{Duration} Projects lasting less than six months have satisfaction rates higher than longer projects.

\subsubsection*{Refelections} Some of the results here presented confirm the intuition and the expectation, for example how \textbf{budget is affecting overall quality} and how \textbf{some Quality Attributes are more relevant for specific business domains}. Other results shows the importance of development methods, other partly contraddict the intuitions, for example project with short duration have 
satisfaction rates. For what reason high satisfaction cannot be achieved and mantained in longer project? While Architectural Styles seems not do significantly affect satisfaction?

This is just a preliminary work and results are synthetized by a reasonbaly large dataset, though it cannot be considered definitive. It is necessary to expand the number of respondents and possibly double it to reach more solid conclusions. On the other side, the nature of domains investigated is pretty broad, therefore to reach some stable conclusion will also be necessary to separately investigate the different domains.

This work is intended to shed some light on the relationships between Quality Attributes and different aspects of software developmnet, and consitutes only a starting point for the accumulation and analysis of further data. In this paper we have indeed estabilished methods and approach to the research which will be applied and extened in future. Next steps are: 

\begin{enumerate}
\item Expand dataset
\item Specialize on business domains
\item Validate the temporary conclusion presented in this work
\end{enumerate}

As a matter of fact, we are already working on the first of these steps.

\bibliographystyle{splncs}
\bibliography{main}

\begin{thebibliography}{10}

\bibitem{ISOIEC9126}
ISO/IEC:
\newblock ISO/IEC 9126. Software engineering -- Product quality.
\newblock ISO/IEC (2001)

\bibitem{6835311}
:
\newblock Ieee standard for software quality assurance processes.
\newblock IEEE Std 730-2014 (Revision of IEEE Std 730-2002) (June 2014)  1--138

\bibitem{Chung}
Chung, L., Nixon, B., Yu, E., Mylopoulos, J.:
\newblock Non-Functional Requirements in Software Engineering.
\newblock International Series in Software Engineering. Springer US (1999)

\bibitem{Garlan}
Shaw, M., Garlan, D.:
\newblock Software Architecture: Perspectives on an Emerging Discipline.
\newblock Prentice-Hall, Inc., Upper Saddle River, NJ, USA (1996)

\bibitem{SA}
Bass, L., Clements, P., Kazman, R.:
\newblock Software Architecture in Practice. 2 edn.
\newblock Addison-Wesley Longman Publishing Co., Inc., Boston, MA, USA (2003)

\bibitem{Hoinville:1982}
Hoinville, G., Jowell, R.:
\newblock Survey Research Practice. 1 edn.
\newblock SCPR (1982)

\bibitem{Silverman:2000}
Silverman, D.:
\newblock Doing Qualitative Research: A Practical Handbook.
\newblock SAGE Publications (2000)

\bibitem{Marshall:1989}
Marshall, C., Rossman, G.:
\newblock Designing Qualitative Research.
\newblock Sage Publications (2006)

\bibitem{methodology1}
Shaughnessy, J., Zechmeister, E., Zechmeister, J.:
\newblock Research Methods in Psychology.
\newblock McGraw-Hill Higher Education. McGraw-Hill (2003)

\bibitem{methodology2}
Groves, R.M., {Fowler, Jr.}, F.J., Couper, M.P., Lepkowski, J.M., Singer, E.,
  Tourangeau, R.:
\newblock Survey Methodology. Second edn.
\newblock Wiley, Hoboken, N.J. (2009)

\bibitem{Kassab}
Kassab, M., Neill, C., Laplante, P.:
\newblock State of practice in requirements engineering: contemporary data.
\newblock Innovations in Systems and Software Engineering \textbf{10}(4) (2014)
   235--241

\bibitem{DeMarco:2002}
DeMarco, T.:
\newblock Slack: Getting Past Burnout, Busywork, and the Myth of Total
  Efficiency.
\newblock Broadway Books (2002)

\end{thebibliography}

%\begin{thebibliography}{5}
%
%\bibitem{Garlan}
%David Garlan, Mary Shaw: Software Architecture: Perspectives on an Emerging Discipline, Prentice Hall (1996)

%\bibitem{SA}
%Bass, L., Clements, P., Kazman, R.: SEI Series in Software Engineering. In: 3rd edition of Software Architecture in Practice, Addison-Wesley Professional (2012)

%\bibitem{DeMarco:2002}
%Tom DeMarco, Slack: Getting Past Burnout, Busywork, and the Myth of Total Efficiency, Crown Business (2002)

%\bibitem{Hoinville:1982} Hoinville, G., R. Jowell: Survey Research Practice, SCPR (1982)

%\bibitem{Marshall:1989} Marshall, C., G.B. Rossman: Designing qualitative research, Sage Publications (1989)

%\bibitem{Silverman:2000} Silverman, D.: Doing qualitative research, Sage Publications (2000)

%\bibitem{methodology1}John Shaughnessy,  Eugene Zechmeister, Jeanne %Zechmeister: 10th edition,  Research methods in psychology, McGraw-Hill Education (2014)

%\bibitem{methodology2}
%Robert M. Groves, Floyd J. Fowler, Jr., Mick P. Couper, James M. Lepkowski, Eleanor Singer, Roger Tourangeau: Survey methodology,WIley publishing (2009)

%\bibitem{Kassab}
%Mohamad Kassab, Colin J. Neill, Phillip A. Laplante: State of practice in requirements engineering: contemporary data. ISSE 10(4): 235-241 (2014)

%\end{thebibliography}

\end{document}